# Motionless and fast measurement technique for obtaining the spectral diffraction efficiencies of a grating

Shenghao Wang, Shijie Liu, a) Jianda Shao, b) Yunxia Jin, Fanyu Kong, and Yonglu Wang

*Key Laboratory of Materials for High Power Laser, Shanghai Institute of Optics and Fine Mechanics, Chinese Academy of Sciences, Shanghai, 201800, China.*

The measurement of the spectral diffraction efficiencies of a diffraction grating is essential for improving the manufacturing technique and for assessing the grating's function in practical applications. The drawback of the currently popular measurement technique is its slow speed due to the hundreds of repetitions of two kinds of time-consuming mechanical movements during the measuring process (i.e., the rotation of the mechanical arm to capture the light beam, and the mechanical variation of the output wavelength of the grating monochromator). This limitation greatly restricts the usage of this technique in dynamic measurement. In this manuscript, we present a motionless and fast measurement technique for obtaining the spectral diffraction efficiencies of a plane grating, effectively eliminating the aforementioned two kinds of mechanical movement. Herein, the proposed solution for removing the first kind of mechanical movement is tested, and the experimental result shows that the proposed method can be successfully used to measure the plane transmission grating's spectral diffraction efficiencies in the wavelength range of 550-750 nm. The method for eliminating the second kind of mechanical movement is not verified in this manuscript; however, we think that it is very straightforward and commercially available. We estimate that the spectral measurement can be achieved on a millisecond timescale by combining the two solutions. Our motionless and fast measuring technique will find broad applications in dynamic measurement environments and mass industrial testing.

## I. INTRODUCTION

Diffraction gratings, currently the most important dispersive element, have been commonly used in the fields of spectral analysis, high-power lasers, optical measurements and many others.[1-5] A diffraction grating's spectral diffraction efficiencies are key parameters when exploring the manufacturing process and evaluating its performance; therefore, the measurement of a diffraction grating's spectral diffraction efficiencies is essential for improving its manufacturing technique and to assess its function in practical applications.[6-9]

$$\eta(\lambda_i) = \frac{I_m(\lambda_i)}{I_0(\lambda_i)}. \qquad (1)$$

Eq. (1) is the definition of a grating's diffraction efficiency, herein expressed as $\eta(\lambda_i)$, at the wavelength of $\lambda_i$ in the $m$ diffraction order. $I_0(\lambda_i)$ and $I_m(\lambda_i)$ represent the power of the incident monochromatic light beam and the diffracted light beam in the $m$ order, respectively. The grating's spectral diffraction efficiencies in the wavelength range of $\lambda_1$ - $\lambda_n$ can be obtained by successively measuring the diffraction efficiency at the wavelengths of $\lambda_1$, $\lambda_2$ ⋯ $\lambda_{n-1}$ and $\lambda_n$.

Over the past half-century, several methods have been introduced to measure the diffraction efficiencies of a grating at particular wavelengths[9-12] or across a continuous wavelength band.[13-15] In 2006, a method based on an optical cavity was introduced to measure a grating with an ultrahigh diffraction efficiency (>99%) in the Littrow configuration,[16,17] and recently, with the aim of increasing measurement accuracy, Fourier spectral technology was used to measure the diffraction efficiency of a plane grating.[18]

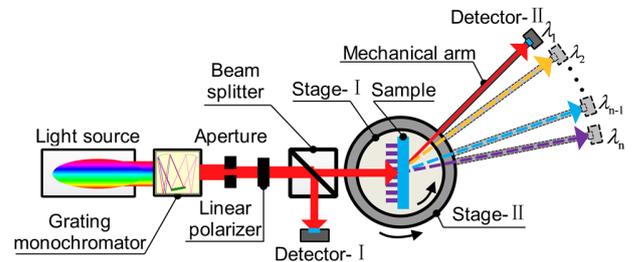

FIG. 1. (Color online). Framework of the system for measuring the spectral diffraction efficiencies of a plane transmission grating using the currently popular method.

An illustration of the method for measuring the spectral diffraction efficiencies of a plane grating that is, to the best of our knowledge, currently the most common is shown in Fig. 1.[13-15] The system is mainly based on a traditional

---

a), b) Authors to whom correspondence should be addressed. Electronic mail: shijieliu@siom.ac.cn,  jdshao@siom.ac.cn.

double-beam photometric framework; it also includes a custom-made mechanical structure containing two motorized rotating stages (stage-I and stage-II) and a mechanical arm that is fixed on stage-II. During the measuring process, the sample is first rotated by Stage-I to the incident angle of $\alpha$, and the wavelength of the light emitted from the grating monochromator is set at $\lambda_1$ (which is realized by mechanically rotating the diffraction grating inside the monochromator). Then, based on the grating diffraction equation in Eq. (2), detector-II rotates to the propagation direction of the grating's diffraction light (expressed as $\beta$) with the help of stage-II and the attached mechanical arm. Here $d$ represents the period of the grating. Subsequently, detector-I and detector-II simultaneously acquire the intensity of the reference and diffracted light beams. Finally, the diffraction efficiency $\eta(\lambda_1)$ at the wavelength of $\lambda_1$ can be obtained by mathematical calculation using the calibration data, which are obtained in the absence of the inspected grating.

$$d \times (\sin\alpha + \sin\beta) = m\lambda. \qquad (2)$$

The output wavelength of the monochromator is then set successively to $\lambda_2$, $\lambda_3$ $\cdots$ $\lambda_{n-1}$ and $\lambda_n$, and at each wavelength, the above procedure is repeated. Thus, the diffraction efficiencies $\eta(\lambda_2)$, $\eta(\lambda_3)$ $\cdots$ $\eta(\lambda_{n-1})$ and $\eta(\lambda_n)$ corresponding to each wavelength can be obtained. Finally, the spectral diffraction efficiencies in the wavelength range from $\lambda_1$ to $\lambda_n$ are available. Note that detector-II needs to rotate to the propagation direction of the grating's diffraction light beam corresponding to the wavelength of $\lambda_i$ at the $i$ th step, based on Eq. (2).

The currently popular measurement technique can be used to automatically measure the spectral diffraction efficiencies of a plane grating with an accuracy of approximately 0.3-0.5%.[15] However, the major drawback of this technique is the slow measuring speed resulting from the hundreds of repetitions of two kinds of time-consuming mechanical movements during the measuring process (the rotation of the mechanical arm to capture the light beam, and the mechanical variation of the output wavelength of the grating monochromator). In fact, the data acquisition is supposed to begin only after the mechanical component is stopped and stabilized. The total time required to obtain the spectral diffraction efficiencies in the expected wavelength range is, therefore, considerably long because of the hundreds of repetitions. For example, approximately 5-8 minutes is needed to obtain the spectral diffraction efficiencies of broadband pulse compression gratings from 700 nm to 900 nm (in sampling steps of 1 nm).[15] The slow measuring speed of the currently popular measurement technique greatly restricts the broad application of this technique in the following circumstances where fast measurement is important. (a) In the manufacturing process of gratings, scatterometry has been widely used as a fast, non-destructive, precise and low-cost way to determine the mean pitch and other dimensional parameters of periodic structures by solving an inverse problem based on the measured diffraction efficiencies.[19-23] In the early days of scatterometry, the diffraction efficiency at a single wavelength was measured at a high speed to monitor on-line the parameters, such as the etch rate in $SiO_2$ and $Si_3N_4$[24] and the line width on a photo mask.[25] Later, by combining the method of rigorous coupled wave analysis, this easy measurement technique was demonstrated to be a powerful tool for in situ characterization.[26-29] However, for robust and fast reconstruction of the scatterometry measurements in some complicated circumstances, many individual data need to be obtained rapidly.[22] The fast measurement of the spectral diffraction efficiencies at many wavelengths was thus considered, but as demonstrated by Eq. (2), obtaining the diffraction efficiency versus wavelength is troublesome and time consuming because the diffraction angles depend strongly on the wavelength of the incident beam. Therefore, in most cases, only the spectral diffraction efficiencies in the zero diffraction order (that is, in the direction of specular reflection) is determined at high speeds with a spectrometer.[30-32] The above techniques have been used very widely in today's commercial state-of-art scatterometers,[33-35] and a very high in situ measurement accuracy is guaranteed. However, considering that the zero order diffraction efficiency is normally not as sensitive as the nonzero orders when changing grating parameters,[35] more details of the grating can therefore be computed (or a certain parameter can be fitted with increasing robustness) if the diffraction efficiencies at many wavelengths (not in the zero diffraction order) can be simultaneously measured in a short time.[22, 35] (b) For many practical applications of diffraction gratings, environmental parameters, such as temperature and humidity, have a non-negligible influence on the microstructure of the grating and, thus, affect the grating's spectral diffraction efficiencies.[36-38] Hence, the dynamic measurement of the grating's spectral diffraction efficiencies in the case, for example, of changing the environmental temperature would provide abundant data on ways to effectively utilize the grating in different environments. (c) In mass industrial manufacturing, the spectral diffraction efficiencies of the grating need to be obtained rapidly to increase the production efficiency.

With the aim of achieving fast measurement of the grating's spectral diffraction efficiencies in a certain wavelength range, we present herein a measurement technique for rapidly obtaining the spectral diffraction efficiencies of a plane grating by eliminating the aforementioned two kinds of mechanical movement. The proposed solution for removing the first kind of mechanical movement is experimentally tested, and the method for eliminating the second kind of mechanical movement is discussed.

## II. MEASUREMENT THEORY OF THE NEW METHOD

The framework of the system for measuring the spectral diffraction efficiencies using the newly proposed method is illustrated in Fig. 2. Compared with the currently popular method (see Fig. 1), the key features of our measuring framework are the employment of a convex lens, an integrating sphere and an acousto-optical tunable filter (AOTF). [39, 40]

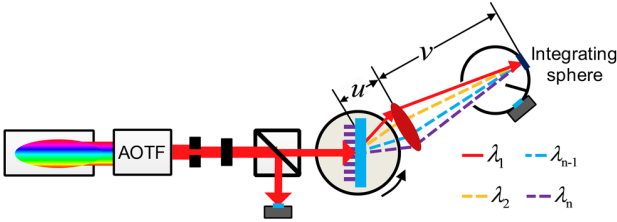

FIG. 2. (Color online). Framework of the system for measuring the spectral diffraction efficiencies of a plane transmission grating using the newly proposed technique.

The convex lens and the integrating sphere are used to remove the first kind of mechanical movement.

$$\frac{1}{u}+\frac{1}{v}=\frac{1}{f}. \quad (3)$$

Based on the mathematical relationship (see Eq. (3)) among the object distance $u$, the image distance $v$ and the focal length $f$ of a convex lens, we know that when changing the wavelength of the incident beam, all the light beams at the same diffraction order diffracted by the grating are focused by the convex lens on one point (strictly speaking, the focal length of the convex lens changes when the wavelength is varied; however, in this situation, the relative variation is very small and can be considered negligible[41]). By mounting the integrating sphere such that the focal point is located on its inner diffuse surface (the diffuse reflectance of the inner surface is almost the same for slight variations in the incident angle[42]), the first kind of mechanical movement can be successfully eliminated.

The use of the AOTF is aimed at eliminating the second kind of mechanical movement. Compared with a grating monochromator (the wavelength tuning of a grating monochromator is achieved by mechanically rotating the inner diffraction grating, which can typically be done on a timescale of sub-seconds or much longer), a key feature of the AOTF is that its output wavelength can be varied electronically at an ultrafast speed (a timescale of a microsecond is available). This feature has been widely used to achieve the dynamic measurement of an absorption spectrum on millisecond timescales.[43, 44]

By combining the two aforementioned solutions, we demonstrate that, compared with the currently popular method, which involves hundreds of repetitions of time-consuming mechanical movement, the proposed method can measure the plane grating's spectral diffraction efficiencies successfully at a high speed without the involvement of mechanical movements (only electronic operations exist in the measurement process).

## III. EXPERIMENTAL SETUP

### A. Layout of the measuring system

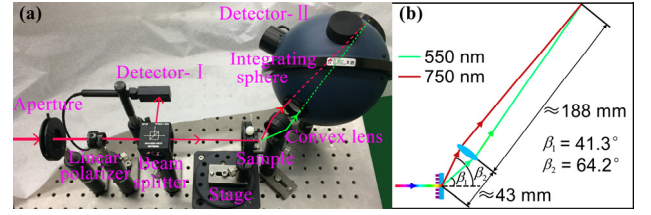

FIG. 3. (Color online). (a) Physical map of the system for measuring the spectral diffraction efficiencies using the newly proposed method, and (b) key geometric dimensions of the system.

A measuring setup is built to verify the proposed solution for removing the first kind of mechanical movement. A photograph of the layout of the system is shown in Fig. 3(a). The system mainly consists of a supercontinuum laser source, a grating monochromator, an aperture, a linear polarizer, an unpolarized beam splitter, a reference detector (detector-I), a motorized rotation stage, a convex lens, an integrating sphere, a test detector (detector-II), a dual channel data acquisition module and a personal computer. The supercontinuum laser source emits a polychromatic spectrum with a total power of 2 W in the wavelength range of 390-2600 nm, and it has a mean power density of approximately 1 mW/nm and a power stability better than ±1%. The grating monochromator can operate from 500 nm to 1200 nm with a wavelength resolution of approximately 0.125 nm, and the spectral full width at half maximum of the generated light beam is approximately 1.0-2.5 nm. When the output wavelength of the monochromator is changed at an increment of 1 nm, a stabilization time of 55 millisecond is needed. The linear polarizer has an effective wavelength range of 550-1500 nm, and its extinction ratio is higher than 1000:1. The unpolarized beam splitter can be applied from 700 nm to 1100 nm with a splitting ratio of approximately 50:50. The reference photoelectric detector is a photodiode-based power sensor with a spectral response wavelength range of 350-1100 nm and a power detection range of between 500 pW and 1 W. The convex lens has a diameter of 25.4 mm, and its focal length is 35 mm. The integrating sphere has an inner diameter of 5.3 inches, and a silicon detector is mounted on the output port; its spectral response wave range is 200-1100 nm, and the power detection range is between 300 nW and 1 W. The data acquisition module is a microprocessor-based, laser power meter containing dual input channels. The sample under inspection is a plane transmission grating working in the visible wavelength range, with a line density of 1200 grooves per mm.

The polychromatic light beam emitted by the supercontinuum laser source is first transmitted by an optical fiber into the grating monochromator, and then, the generated monochromatic beam passes successively through the optical aperture and the linear polarizer. Subsequently, reference and test light beams are simultaneously generated by the unpolarized beam splitter; detector-I collects the reference light beam, while the integrating sphere and detector-II capture the test light beam after it is diffracted by the grating and then deflected by the convex lens. In this way, the negative effect of laser power fluctuation on measurement accuracy can be successfully eliminated by the adoption of a double-beam configuration. The photon signal detected by the two detectors is first photoelectrically converted and then digitized by the dual channel data acquisition module. The personal computer is used to remotely control the grating monochromator and the motorized rotation stage and to operate the data acquisition module through serial ports. LabVIEW-based software has been developed to perform functions such as wavelength setting and scanning, sample rotation, data acquisition and post-processing.

Fig. 3(b) represents the key geometric dimensions of the measuring system. The plane transmission grating is measured at normal incidence. The distance between the sampling point in the grating and the convex lens is approximately 43 mm, and the integrating sphere's inner surface is approximately 188 mm beyond the convex lens. The wavelength range measured here is 550-750 nm with a sampling step of 1 nm at the -1 diffraction order. The red and green lines show the propagation paths of the diffracted light beams when the wavelength of the incident light is 550 nm and 750 nm, respectively, and the diffraction angles of the plane transmission grating are 41.3° and 64.2°, respectively.

**B. Testing process and working theory**

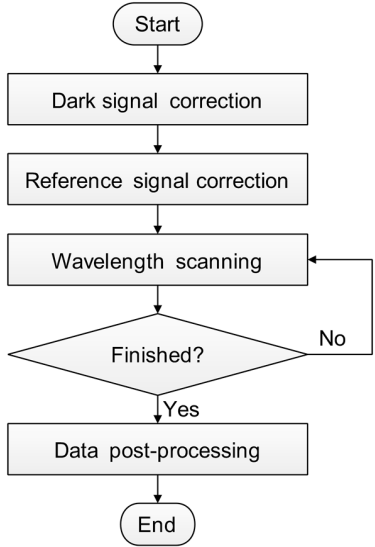

FIG. 4. Flow chart of the testing procedure.

Fig. 4 is a flow chart of the testing procedure of the proposed method; we describe it in detail below.

S1: Simultaneously capture the dark signals of detector-I and detector-II, and record them as $I^1_{dark}$ and $I^2_{dark}$, respectively.

S2: Mount the convex lens in the test beam path between the unpolarized beam splitter and the integrating sphere; set the output wavelength of the monochromator as $\lambda_1$; and then, let the integrating sphere directly collect the test light beam without penetrating through the sample.

S3: Simultaneously capture the electric signals of detector-I and detector-II; record them as $I^1_{ref}(\lambda_1)$ and $I^2_{ref}(\lambda_1)$, respectively; and compute the intensity ratio $k(\lambda_1)$ as follows:

$$k(\lambda_1) = \frac{I^2_{ref}(\lambda_1) - I^2_{dark}}{I^1_{ref}(\lambda_1) - I^1_{dark}}. \quad (4)$$

S4: Successively set the output wavelength of the monochromator as $\lambda_2$, $\lambda_3$ ⋯ $\lambda_{n-1}$ and $\lambda_n$ while repeating S3 at each individual wavelength, and compute $k(\lambda_2)$, $k(\lambda_3)$ ⋯ and $k(\lambda_n)$.

S5: Mount the grating on the motorized rotation stage; move the convex lens and the integrating sphere to the position shown in Fig. 3(b); and then, set the output wavelength of the grating monochromator as $\lambda_1$.

S6: Acquire the intensity of the reference and diffraction light beams simultaneously; record them as $I^1_{sig}(\lambda_1)$ and $I^2_{sig}(\lambda_1)$, respectively; and compute the intensity ratio $k^*(\lambda_1)$ as follows:

$$k^*(\lambda_1) = \frac{I^2_{sig}(\lambda_1) - I^2_{dark}}{I^1_{sig}(\lambda_1) - I^1_{dark}}. \quad (5)$$

S7: Successively set the output wavelength of the monochromator as $\lambda_2$, $\lambda_3$ ⋯ $\lambda_{n-1}$ and $\lambda_n$ while repeating S6 at each individual wavelength, and compute the intensity ratios $k^*(\lambda_2)$, $k^*(\lambda_3)$ ⋯ $k^*(\lambda_{n-1})$ and $k^*(\lambda_n)$.

S8: Use Eq. (6) to calculate the diffraction efficiencies $\eta(\lambda_1)$, $\eta(\lambda_2)$, $\eta(\lambda_3)$ ⋯ $\eta(\lambda_{n-1})$ and $\eta(\lambda_n)$ of the sample corresponding to the wavelengths of $\lambda_1$, $\lambda_2$, $\lambda_3$ ⋯ $\lambda_{n-1}$ and $\lambda_n$.

$$\eta(\lambda_i) = \frac{k^*(\lambda_i)}{k(\lambda_i)}. \quad (6)$$

**IV. EXPERIMENTAL RESULTS**

Fig. 5(a) and Fig. 5(b) show the curves of the intensity ratio vs. wavelength results measured using the new measurement system and the computed spectral diffraction efficiencies in the cases of P- and S-polarization, respectively. In Fig. 5(a) and Fig. 5(b), the black curves demonstrate the relation of the intensity ratio vs. wavelength without a sample in the beam path, the red curves are the intensity ratio vs. wavelength with a sample in the beam path, and the blue curves depict the computed spectral diffraction efficiencies.

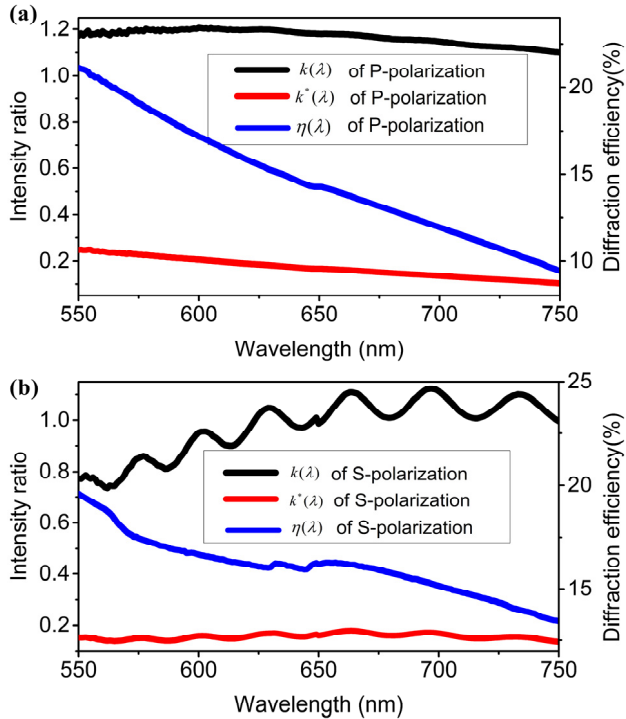

FIG. 5. (Color online). (a) and (b) show the measured curves of intensity ratio vs. wavelength and the computed spectral diffraction efficiencies in the cases of P- and S-polarization, respectively. Black curves: intensity ratio vs. wavelength without a sample in the beam path; red curves: intensity ratio vs. wavelength with a sample in the beam path; blue curves: spectral diffraction efficiencies. An artifact exists at approximately 650 nm (see the black curve in (b)) and is caused by the switch of the diffraction grating inside the monochromator.

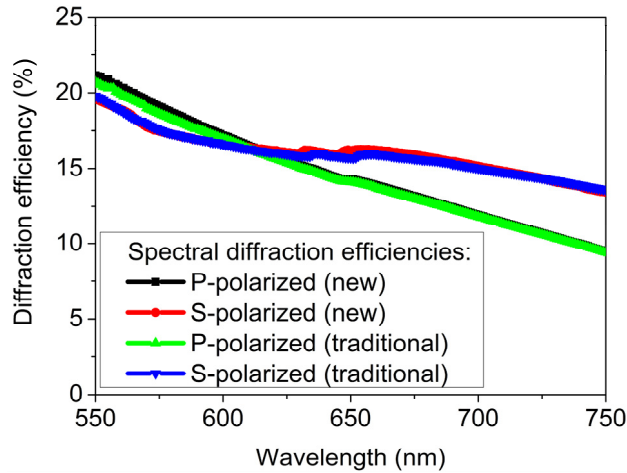

FIG. 6. (Color online). Black, red, green and blue curves are the measured spectral diffraction efficiencies of, respectively, the P-polarized light using the new method, the S-polarized light using the new method, the P-polarized light using the currently popular method and the S-polarized light using the currently popular method.

Fig. 6 shows the obtained spectral diffraction efficiencies of the inspected plane transmission grating at the -1 diffraction order in the wavelength range of 550-750 nm. The spectral diffraction efficiencies of P- and S-polarization at normal incidence are measured using the currently popular setup (as shown in Fig. 1) and the newly upgraded system (see Fig. 2 and Fig. 3). In Fig. 6, the black and green curves represent the spectral diffraction efficiencies of the P-polarized light measured using the new and currently popular methods, respectively. Accordingly, the red and blue curves show the data obtained when the grating is illuminated by S-polarized light employing the two measuring frameworks. Note that both the currently popular and newly proposed measurement systems have been calibrated by a standard sample with known optical transmittance before the aforementioned measurements.

The relative error between the measured spectral diffraction efficiencies using the new and currently popular measuring methods is 1.44% in the case of P-polarization and 1.13% in the case of S-polarization. The relative error is computed using the following equation:

$$\delta = \sqrt{\frac{\sum_{i=550}^{750}\left[\frac{\eta_{new}(\lambda_i) - \eta_{trad}(\lambda_i)}{\eta_{trad}(\lambda_i)}\right]^2}{750-550}}, \quad (7)$$

where $\eta_{new}(\lambda_i)$ and $\eta_{trad}(\lambda_i)$ represent the diffraction efficiencies at the wavelength of $\lambda_i$ measured by the new and currently popular measuring frameworks, respectively.

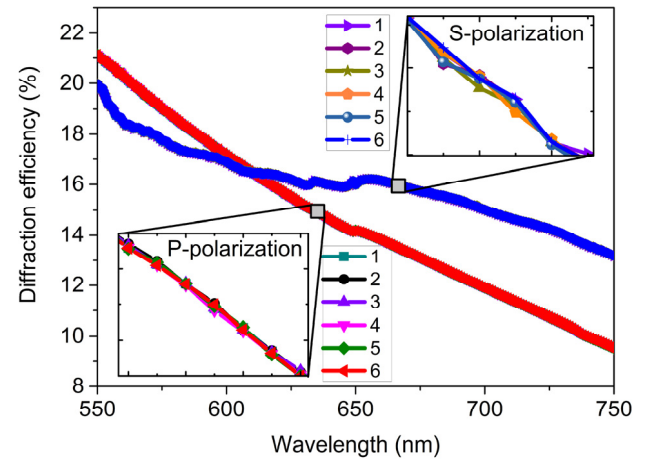

FIG. 7. (Color online). Six repeated measuring results. Left bottom: the enlarged spectral diffraction efficiencies of the P-polarized light; (b) top right: the enlarged spectral diffraction efficiencies of the S-polarization.

Shown in Fig. 7 are the 6 repeated measuring results obtained using the new method in the cases of P- and S-polarization at normal incidence. The overlapping curves in red and blue demonstrate the 6 measured spectral diffraction efficiencies of P- and S-polarized light; see the enlarged view in the left bottom and top right for details. The repeatability errors of the curves for P- and S-polarization are 0.043% and 0.046%, respectively, which were determined with the following formula.

$$\zeta = \frac{\sum_{k=1}^{201} \chi_{\lambda_k}}{201}, \quad (8)$$

where $\chi_{\lambda_k}$ is the repeatability error at the wavelength of $\lambda_k$. It was computed as follows:

$$\chi_{\lambda_k} = \frac{\sqrt{\sum_{i=1}^{6}\left(\eta_i - \sum_{i=1}^{6}\eta_i/6\right)^2/6}}{\sum_{i=1}^{6}\eta_i/6}. \quad (9)$$

In Eq. (9), $\eta_i$ represents the diffraction efficiency of the plane transmission grating at the wavelength of $\lambda_k$ in the $i$ th measurement.

## V. DISCUSSION

Based on the comparison of the measured spectral diffraction efficiencies shown in Fig. 6 and the computed relative errors between the obtained spectra, we can conclude that the proposed solution for removing the first kind of mechanical movement has considerably good feasibility.

TABLE I. Absolute measurement error resulting from the fact that the convex lens's optical transmittance is a function of AOI in the aforementioned measurement.

| Wavelength & polarization | | Absolute error |
|---|---|---|
| 550 nm | S-polarization | 0.616% |
| | P-polarization | 0.511% |
| 750 nm | S-polarization | 0.411% |
| | P-polarization | 0.239% |

Here, it should be noted that one obvious error source of the new measurement method is that the convex lens's optical transmittance is a function of the angle of incidence (AOI). In the measurement procedure, i.e., S2-S4, all the reference signals, that is $k(\lambda_1)$, $k(\lambda_2)$ ⋯ $k(\lambda_n)$, were captured while the laser beam propagated along the convex lens's optical axis in the absence of the sample. However, in S6 and S7, after inserting the grating into the beam path, $k^*(\lambda_1)$, $k^*(\lambda_2)$ ⋯ $k^*(\lambda_n)$ were obtained after the laser beam was first diffracted by the grating and then penetrated through the convex lens. Consequently, normal incidence cannot be entirely guaranteed (see the red and green beam paths in Fig. 3). The transmittance difference between the two cases clearly introduces some degree of measurement error. By theoretically calculating the convex lens's transmittance at different wavelengths, polarizations and AOI in our experimental configuration based on the Fresnel equations (the optical absorption in the lens was also taken into account), this measurement errors can be evaluated. Table I lists the absolute measurement errors resulting from this error source at the wavelengths of 550 nm and 750 nm (we think that the measurement error is greatest when the AOI difference is the largest). The measurement error in our presented measurements is acceptable, and this error can be successfully eliminated by theoretical calculation or experimental testing. Furthermore, applying a coating of anti-reflection film on the convex lens is another way to remove this error.

The 6 repeated measuring results shown in Fig. 7 show that measuring accuracy and repeatability are very high using the new measurement method. In fact, the high accuracy of repeated measurements is another advantage of the new method over the currently popular one, in which the mechanical movement inevitably introduces error into repeated measurements (because the spectral responsivity is not perfectly uniform in the photosensitive area of the detector, mechanical movement would lead to variation in the voltage signals generated by the photo detector).

To measure the spectral diffraction efficiencies of the studied plane grating in the wavelength range of 550-750 nm with a sampling step of 1 nm, approximately 400 seconds were needed when adopting the currently popular method. However, in the newly developed system, the elimination of only one kind of mechanical movement does not markedly increase the measuring speed. The process is still hindered by the second kind of mechanical movement and the currently used data acquisition software. To achieve ultrafast measurements, a real-time operating system, such as a field programmable gate array, is needed to control the instruments and collect the data.

TABLE II. Comparison of the time consumed between the measurements of the spectral diffraction efficiencies using the new and currently popular methods (in the case of obtaining the spectral diffraction efficiencies from 700 nm to 900 nm with an interval of 1 nm).

| Measuring method | Time consumed |
|---|---|
| The currently popular method | ≈ 400 seconds |
| The new method | Tens of milliseconds (estimated) |

The adoption of an AOTF to eliminate the second kind of mechanical movement is not experimentally tested in this manuscript due to the current hardware configuration of our laboratory. However, considering that AOTFs have been used widely to attain ultrafast measurement of an absorption spectrum on millisecond timescales, this timescale should be the total time needed to realize hundreds of repetitions of wavelength scanning and data acquisition,[43, 44] we think that our experimental results fully demonstrate the feasibility of our newly proposed motionless and fast measurement technique, and Table II shows the time consumed for the measurements of the spectral diffraction efficiencies using both the new (estimated) and currently popular methods.

Finally, it should be noted that one shortcoming of the proposed measurement technique is the troublesome arrangement of the system for a given grating in a certain wavelength range (see step S5 in the testing process). Additionally, the spectral resolution of an AOTF (a resolution of approximately 0.5 nm is typically considered very high in the visible light range[45]) is usually not as high as that of a grating monochromator.

## VI. CONCLUSION

In conclusion, we present a motionless and fast measurement technique for obtaining the spectral diffraction efficiencies of a plane grating, and its feasibility is tested and confirmed by experimental results. Compared with the currently popular technique, the advantage of the proposed method is that the measurement speed can be greatly increased. Fast time-resolved measurement of the grating's spectral diffraction efficiencies in a certain wavelength range is, therefore, promising and will find broad applications in dynamic measurement and in mass industrial testing.

## ACKNOWLEDGMENTS

This research was supported by the National Natural Science Foundation of China (Nos. 61705246 and 11602280). The authors thank the anonymous reviewers for their constructive comments, which have improved this manuscript.